\documentclass[conference]{IEEEtran}
\usepackage{url}
\usepackage{algorithm}
\usepackage{algorithmic}
\usepackage{ upgreek }
\ifCLASSINFOpdf
\usepackage[pdftex]{graphicx}
\else
\usepackage[dvips]{graphicx}
\fi
\usepackage{amsmath}
\usepackage{color}
\usepackage{ dsfont }
\usepackage[misc]{ifsym}
\usepackage{algorithmic}
\usepackage{array}
\usepackage{verbatim}
\ifCLASSOPTIONcompsoc
\usepackage[caption=false,font=normalsize,labelfont=sf,textfont=sf]{subfig}
\else
\usepackage[caption=false,font=footnotesize]{subfig}
\fi
\usepackage{amsmath}

\begin{document}

\title{Optimal Storage Control for Dynamic Pricing}

\author{\IEEEauthorblockN{Jiaman Wu, Zhiqi Wang, Yang Yu, \emph{and} Chenye Wu$^\text{\Letter}$}
\IEEEauthorblockA{Institute for Interdisciplinary Information Sciences\\
Tsinghua University, Beijing, 100084, P.R. China\\
Email:chenyewu@tsinghua.edu.cn}
}

\maketitle

\begin{abstract}
Renewable energy brings huge uncertainties to the power system, which challenges the traditional power system operation with limited flexible resources. One promising solution is to introduce dynamic pricing to more consumers, which, if designed properly, could enable an active demand side. To further exploit flexibility, in this work, we seek to advice the consumers an optimal online control policy to utilize their storage devices facing dynamic pricing. Towards designing a more adaptive control policy, we devise a data-driven approach to estimating the price distribution. Simulation studies verify the optimality of our proposed schemes.\footnote{This work has been supported in part by Turing AI Institute of Nanjing, and Zhongguancun Haihua Institute for Frontier Information Techonology.}
\end{abstract}

\vspace{0.1cm}


\IEEEpeerreviewmaketitle
\section{Introduction}
A high penetration of renewables in power system is expected to reduce greenhouse gas emissions over the next few years. Meanwhile, the highly stochastic nature of renewables calls for a new paradigm of power system control. In contrast to the traditional paradigm, the most important feature in the new one is more flexibility. One way to recruit additional flexibility is to design proper dynamic pricing schemes to enable active demand side management \cite{khan2016load}.

\subsection{Challenges and Opportunities}
The classical impediment for dynamic pricing comes from the technological difficulty to ensure real time communication. Such impediment is diminishing with improved telecommunication system as well as the widely deployed smart meter devices \cite{borenstein2002dynamic}. 
The other hurdle that prevents dynamic pricing from wide implementation is the public concern over price volatility. In fact, it offers both risks and potential benefits to the consumers. With the decreasing cost of storage system, the consumers can utilize storage system with proper control policy, which can help the consumers by hedging against the risk in dynamic pricing and help the power system by providing more flexibility. In this paper, we seek to design such a policy and achieve the two goals simultaneously.

\begin{figure}[t]
\centering
\includegraphics[width=2.7in]{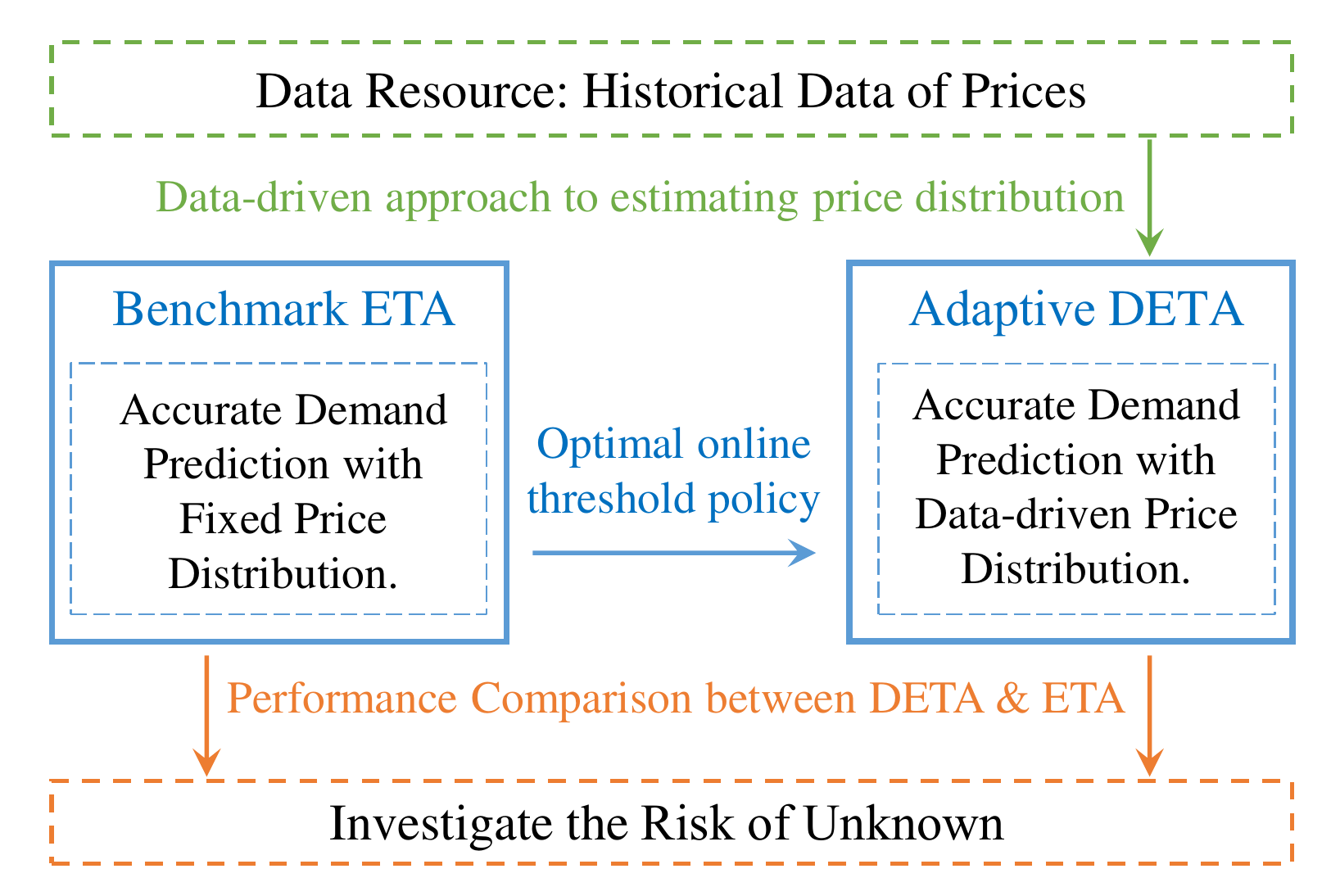}
\caption{Our paradigm for optimal storage control.}\vspace{-0.5cm}
\label{fig 1}
\end{figure}

\subsection{Literature Review}
The investigation on storage control policy for dynamic pricing only emerges recently. Jin \textit{et al.} propose a heuristic algorithm using Mixed Integer Linear Programming to optimize the electric vehicle chagring schedules in \cite{6461500}. Oudalov \textit{et al.} focus on conducting peak load shaving and introduce a sizing methods as well as an optimal operational scheme in \cite{4538388oudalov}. Wang \textit{et al.} design an optimal control policy and solve the optimal investment problem for general ToU scheme using dynamic programming in \cite{wang2020storage}. Chau \textit{et al.} assume the knowledge of future demand and the bounds of prices, and illustrate a threshold cost minimizing online algorithm with worst-case performance guarantee in \cite{chau2016cost}. To deal with limited information and uncertainty, Qin \textit{et al.} introduce an online modified greedy algorithm for storage control in \cite{qin2015online}. Vojvodic \textit{et al.} design a forward threshold algorithm to manage storage operation in real-time market, where stages are decomposed using integer programming and heuristic search  \cite{vojvodic2016forward}. 

In contrast to the literature, we propose an optimal online threshold policy for storage control. Specifically, to highlight the impact of uncertainties in dynamic price on the control policy design, we assume the consumer's own demand prediction is accurate in the near future. This is reasonable as on the grid level, the impact of uncertainties in renewables is reflected through the volatility in dynamic pricing. As shown in Fig. \ref{fig 1}, we design two control policies: one with the accurate price distribution information, the other with the data-driven price distribution estimator. Comparing these two policies reveals the risk of unknown in storage control facing dynamic pricing. 

\subsection{Our Contribution}
To better contrast our work from the literature, we highlight our contributions as follows:
\begin{itemize}
    \item \textit{Optimal Online Threshold Policy}: Based on the one-shot load decomposition technique, we seek to solve the optimal online storage control problem. We show that assuming perfect knowledge of consumer's demand and exact price distribution, a simple threshold policy (Expected Threshold Algorithm, ETA) can minimize the consumer's expected electricity bill.
    \item \textit{Data-driven Distribution Estimator}: We relax the assumption of knowledge on exact price distribution. Instead, when only knowing the type of price distribution, we design a data-driven distribution estimator, yielding Data-driven Expected Threshold Algorithm (DETA).
    \item \textit{Risk of Unknown}: By comparing the performance of two proposed algorithms (ETA and DETA) with the offline optimal, as well as observing the algorithms' performance with accumulating data, we evaluate the cost for risk of unknown through numerical studies.
\end{itemize}

The rest of our paper is organized as follows. Section \ref{sec:System} introduces the system model, and revisits the one-shot load decomposition technique. We propose the optimal online threshold storage control policy in Section \ref{sec: EAT}. Section \ref{sec: DEAT} derives the data-driven distribution estimator for a more adaptive control policy. We evaluate the performance of proposed control policies through simulation studies in Section \ref{sec:SimulationStudy}. Concluding remarks are given in Section \ref{sec:con}.

\begin{figure}[t]
\centering
\includegraphics[width=2.8in]{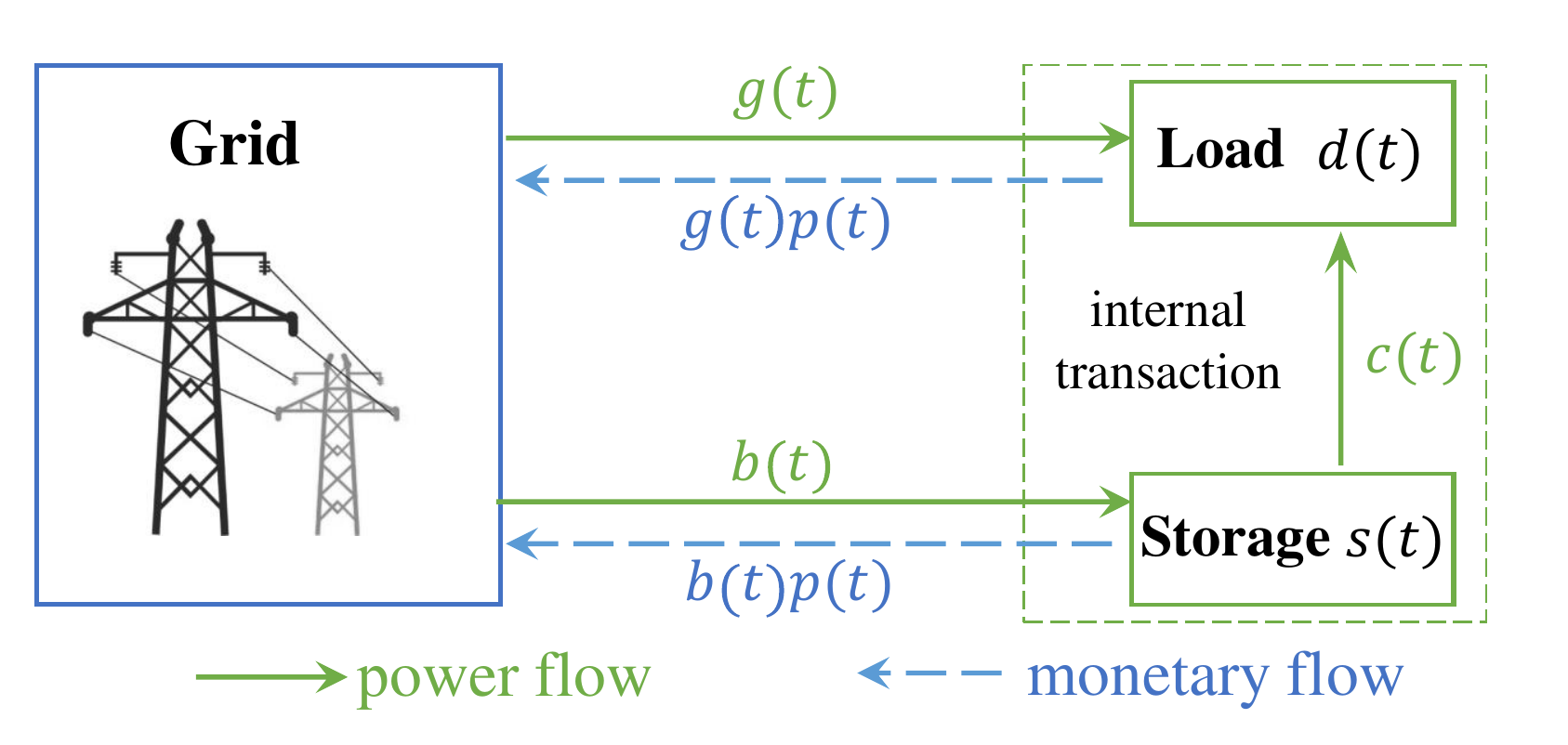}
\DeclareGraphicsExtensions.
\caption{Demonstration of system model.}\vspace{-0.5cm}
\label{fig 2}
\end{figure}

\section{System Model}
\label{sec:System}
Consider the interaction between consumers and the grid as shown in Fig. \ref{fig 2}. The grid operator sets dynamic price $p(t)$ at each time $t$. Facing such a pricing scheme, the consumer wants to satisfy its demand $d(t)$ in different ways: directly purchase energy $g(t)$ from the grid, save energy $b(t)$ in the storage system, or use the energy in the storage system ($c(t)$ out of $s(t)$ in the storage) to meet its demand.

As discussed in the previous section, we assume the demand prediction for each consumer is rather accurate while all uncertainties in the system are reflected in the dynamic pricing's volatility. Even with this assumption, the decision making for each consumer is still quite challenging due to the uncertainty in the future prices and the physical constraints (capacity constraints) coupling all the storage control decisions.

Inspired by \cite{chau2016cost}, in this section, we first revisit the one-shot load decomposition technique, which allows us to decouple the storage control actions across time. Then, we formally define our one-shot load serving problem.

\begin{figure}[t]
\centering
\includegraphics[width=2.7in]{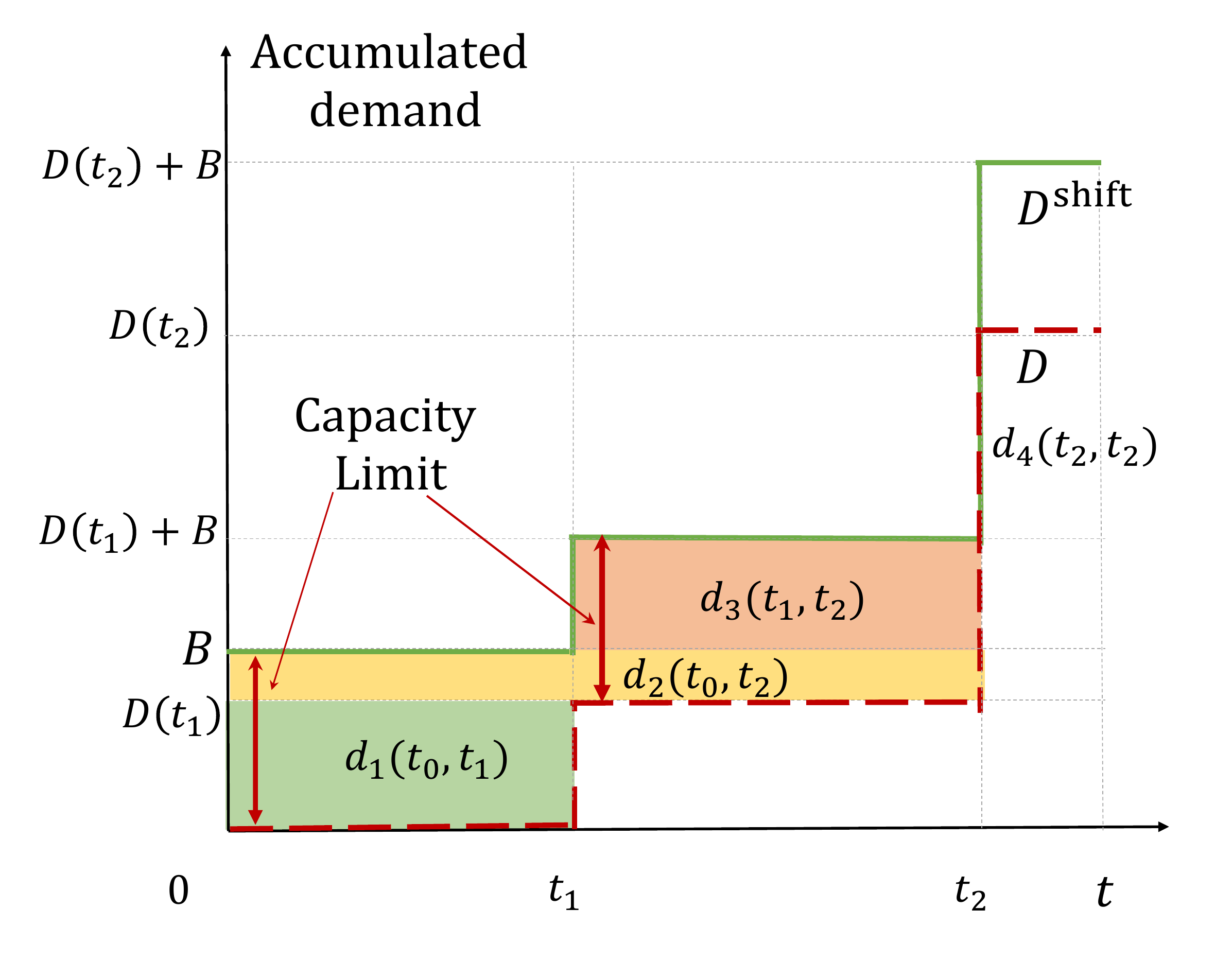}
\DeclareGraphicsExtensions.
\caption{Example for one-shot load decomposition.}\vspace{-0.5cm}
\label{fig 3}
\end{figure}

\subsection{Revisit the One-shot Load Decomposition}
The one-shot load decomposition technique was first proposed by Chau \textit{et al.} in \cite{chau2016cost} to decouple the storage capacity constrained optimization problem into a sequence of one-shot load serving problems.

The idea is simple. Suppose we need serve an accumulated demand as shown by the red dashed curve in Fig. \ref{fig 3}: serving a load of $D(t_{1})$ at time $t_{1}$, and serving a load of $D(t_{2})-D(t_{1})$ at time $t_{2}$. If there were no storage devices, then we have to purchase the load at the time of serving, tolerating all the price volatility. However, with a storage of capacity $B$, we have the choice of purchasing the demand over the whole time span. The one-shot load decomposition technique is designed to highlight such flexibility in serving load. The difficulty is again due to the storage capacity constraint.

Back to our example in Fig. \ref{fig 3}, assume
\begin{equation}
B>D(t_{1}):= d_{1}(t_{0},t_{1}),
\end{equation}
then it is straightforward to see that with storage, we can serve $D(t_{1})$
at any time between $[t_{0},t_{1}]$. Therefore, we define $d_{1}(t_{0},t_{1})=D(t_{1})$, which highlights its flexibility in the time span $[t_{0},t_{1}]$. Next, we assume
\begin{equation}
B<D(t_{2})-D(t_{1}),
\end{equation}
then the demand to be met at $t_{2}$ need be decomposed into three different kinds of demand: $d_{2}(t_{0},t_{2})$, $d_{3}(t_{1},t_{2})$, and $d_{4}(t_{2},t_{2})$. Since $B>D(t_{1})$, the storage has certain spare capacity to store energy and serve the load at $t_{2}$ even between $[0,t_{1}]$ (We assume the perfect knowledge of load). This observation leads to the first kind of demand $d_{2}(t_{0},t_{2})$, which is reserved for load at $t_{2}$ between $[0,t_{2}]$ (the union of $[0,t_{1}]$ and $[t_{1},t_{2}]$). The second kind of demand is due to the released capacity after serving load at $t_{1}$, which is only flexible after $t_1$, which is denoted by $d_{3}(t_{1},t_{2})$ in our example. Note that $B<D(t_{2})-D(t_{1})$, certain amount of load at time $t_{2}$, $d_{4}(t_{2},t_{2})$ has to be met in real time. This leads to the last type of decomposed demand in the one-shot load decomposition technique. 

\vspace{0.1cm}
\noindent \textbf{Remark}: Note that the last type of decomposed demand naturally decouples the original optimization problem over a \emph{long time} span into a set of \emph{smaller scale} optimization problems. This implies that when we conduct the one-shot load decomposition, we only need to look ahead for a couple of hours, the load prediction of which can be rather accurate. This also illustrates our assumption of perfect knowledge on the near future demand is realistic.

Next, we formally introduce the construction process for the one-shot load decomposition technique as follows \cite{chau2016cost}:
\begin{enumerate}
    \item Define the accumulative demand curve $D(t)$.
    \item Define the upward shift accumulative demand curve $D^{\text{shift}}(t)$, which is obtained by shift $D(t)$ by $B$.
    \item Obtain one-shot demand $d_{i}(t_{s}^{i},t_{e}^{i})$ through sandwiched rectangle $(t_{s}^{i}-t_{e}^{i})d_{i}(t_{s}^{i},t_{e}^{i})$ between $D(t)$ and $D^{\text{shift}}$(t).
\end{enumerate}

\subsection{One-shot Load Serving Problem}
We can now focus on the one-shot load serving problem. For each decomposed load $d_{i}(t_{s}^{i},t_{e}^{i})$, which need be served between $t_{s}^{i}$ and $t_{e}^{i}$, we seek to find the time slot $t$ with the minimum price $p(t)$ for serving. In fact, it suffices to understand the stylized one-shot load serving problem, where consumer need satisfy its \textit{one-unit} demand between $0$ and $T$.

Mathematically, the consumer makes a sequence of decisions $u(0),...,u(T)$, where $u(t)$ denotes its purchasing decision at time $t$: if $u(t)=0$, the consumer won't purchase anything; if $u(t)=1$, the consumer meets the unit demand:
\begin{alignat}{2}
\min_{u(0),...,u(T)}\quad & \sum\nolimits_{t=0}^{T}u(t)p(t)\\
\mbox{s.t.}\quad
&u(t)\in \{0,1\}, &\quad & 0\leq t \leq T\\
&\sum\nolimits_{t=0}^{T} u(t) = 1.
\end{alignat}
If we were able to solve the problem in an offline manner, this optimization problem is simply to select:
\begin{equation}
    t^{*}=\mathop{\arg\min}\limits_{t\in[0,T]}p(t).
\end{equation}

However, we cannot foresee the future prices, which warrants designing an online algorithm to solve the one-shot load serving problem. To simplify our subsequent analysis with more insights, we make the following assumption:

\vspace{0.1cm}
\noindent\emph{Assumption}: Dynamic price $p(t)$'s are i.i.d random variables.

\vspace{0.1cm}

With this assumption, we analyze three scenarios: 

\begin{itemize}
    \item \textit{Perfect Prediction}: We use the offline optimal to serve as the benchmark for comparison.
    \item \textit{Exact Distribution}: Knowing the exact distribution (exact parameters) of the random price $p(t)$, we seek to design the optimal online control policy.
    \item \textit{Type of Distribution}: In this scenario, we only know the type of distribution that random price $p(t)$ follows. We devise the data-driven distribution estimator, and incorporate this estimator into the online policy.
\end{itemize}

\section{Optimal Control with Exact Distribution}
\label{sec: EAT}

Based on the one-shot load decomposition, we derive the optimal online storage control policy in this section. We assume that we know the exact price distribution for $p(t)$'s.

For the one-shot load serving problem between $[0,T]$, at each time slot $t$, we can only make two decisions: to purchase the unit demand or not. The two decisions correspond to different expected costs: $p(t)$ for purchasing and $\mathds{E}[w_{t+1}]$ for not purchasing, where $\mathds{E}[w_{t+1}]$ denotes the expected cost for the one-shot unit load serving between $[t+1,T]$.

\subsection{Control Policy ETA}
To characterize this binary choice, we employ a threshold $\theta(t)$ to balance the expected costs between two actions. Hence, the optimal threshold would require
\begin{equation}
\theta(t)=\mathds{E}[w_{t+1}],
\end{equation}
if $p(t) \leq \theta(t)$, we choose to purchase the unit load at time $t$. Otherwise, we defer this action to later time slots.

\begin{figure*}[!t]
\centering
\subfloat[Uniform distribution]{\includegraphics[width=2in]{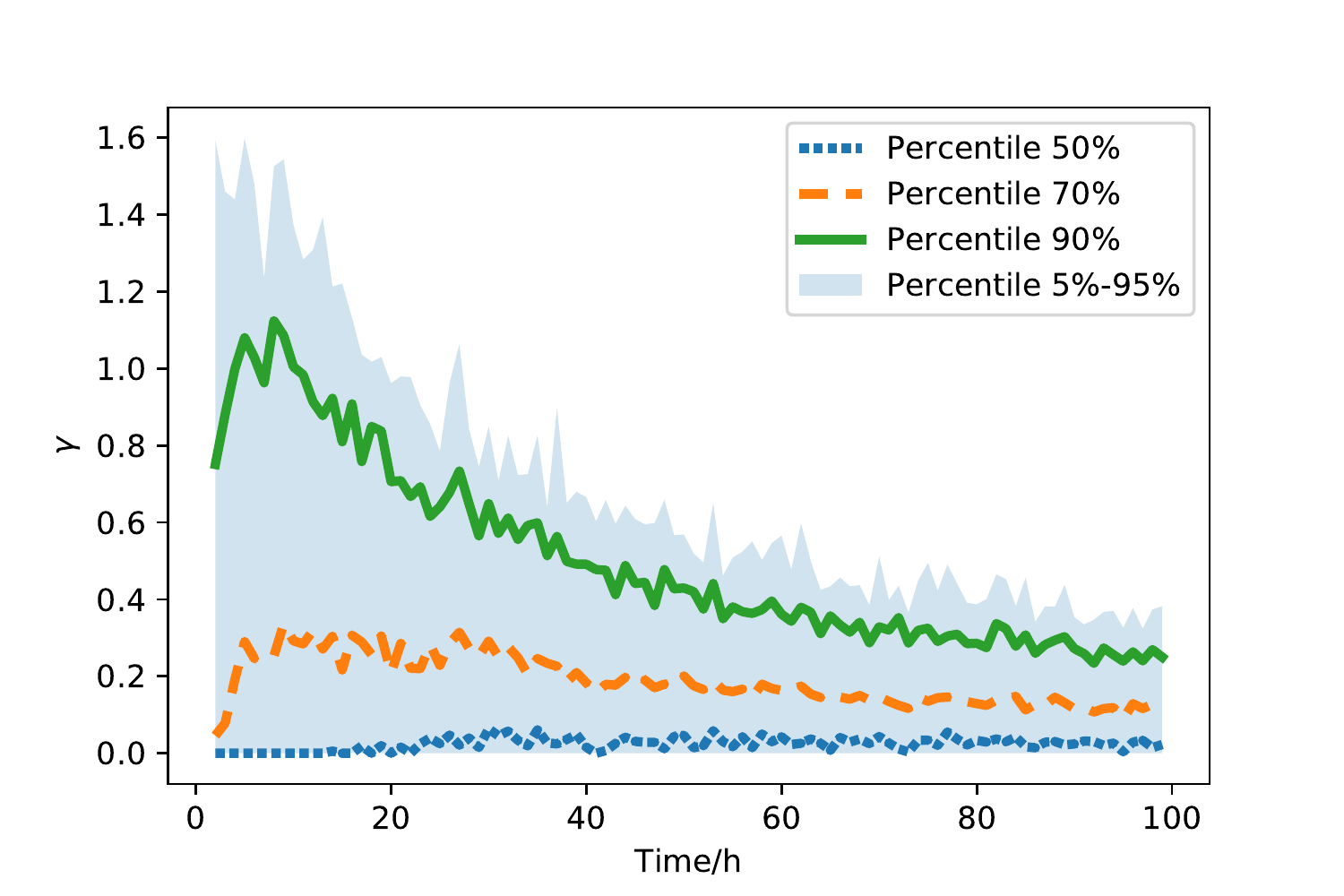}}
\hfil
\subfloat[Half-normal distribution]{\includegraphics[width=2in]{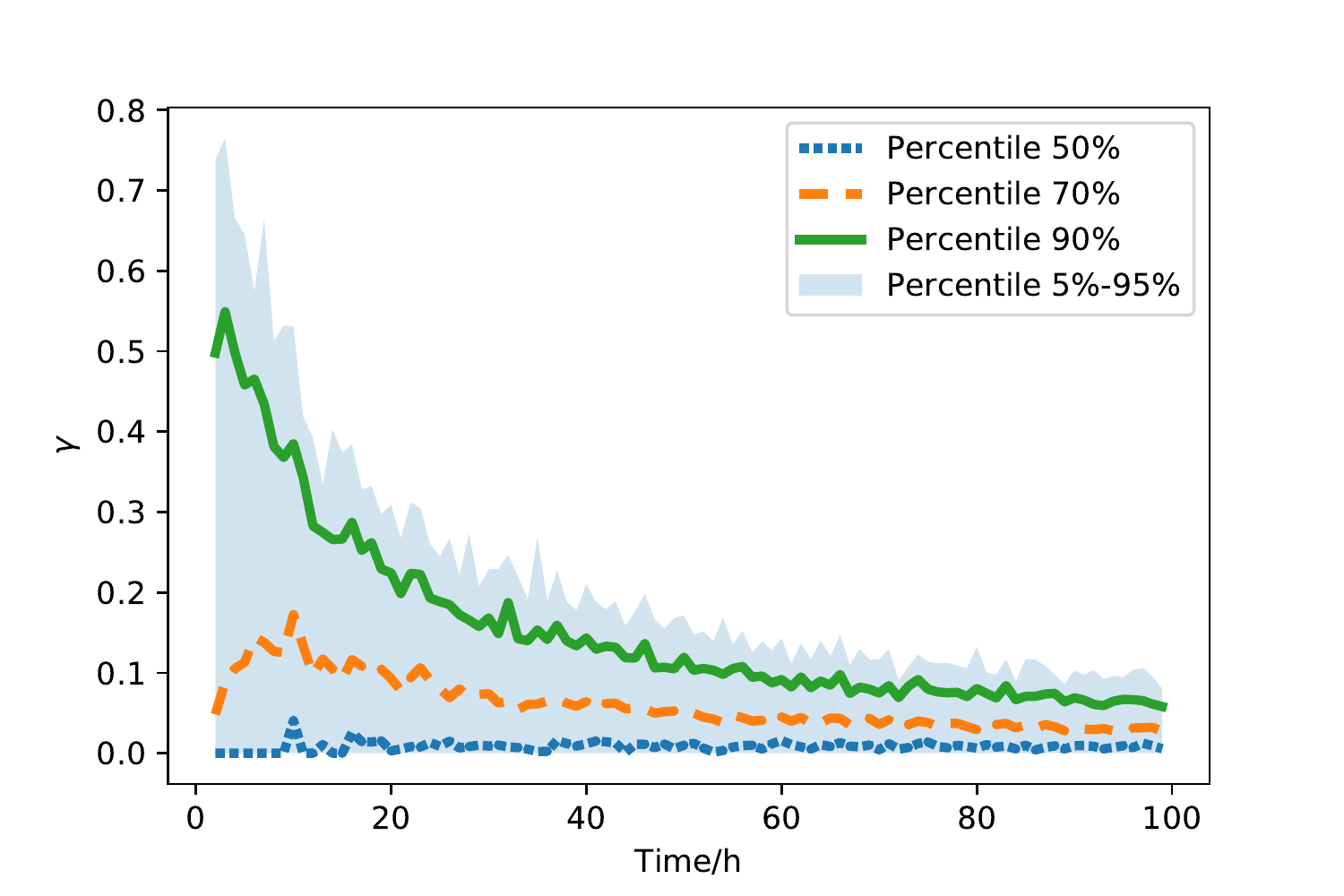}}
\hfil
\subfloat[Log-normal distribution]{\includegraphics[width=2in]{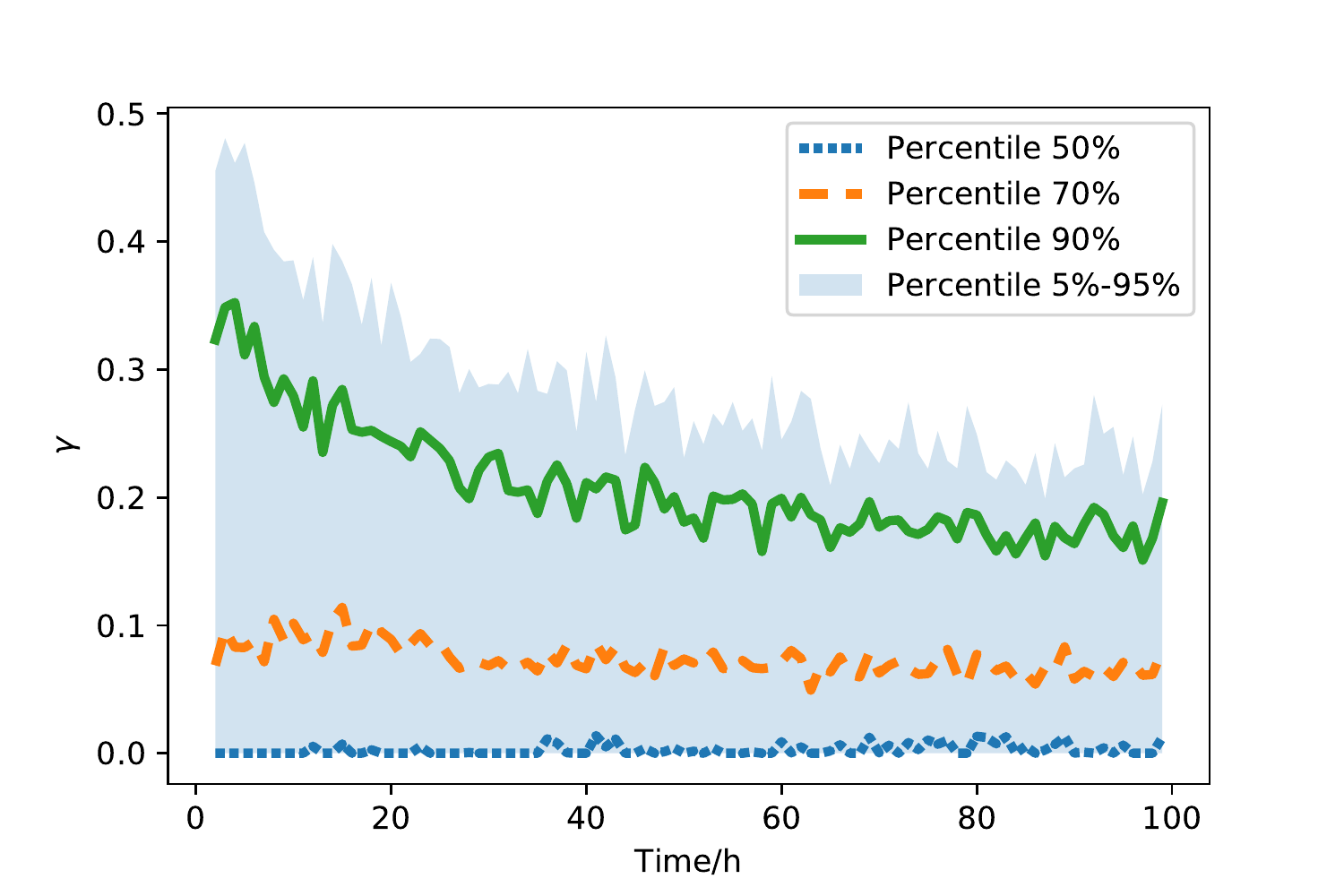}}
\caption{Regret ratio of one-shot serving problem.}\vspace{-0.3cm}
\label{fig 5}
\end{figure*}

Note that the expected cost $\mathds{E}[w_{t}]$'s can be obtained in an recursive manner:
\begin{equation}
\mathds{E}[w_{t}]=\int_{0}^{\theta_{t}}xf_{p(t)}(x)dx+\int_{\theta_{t}}^{\infty}\mathds{E}[w_{t+1}]f_{p(t)}(x)dx,
\label{star}
\end{equation}
where $f_{p(t)}$ is the probability density function of the dynamic price $p(t)$. Hence, our desirable decision making $u(0),u(1),...,u(t)$ can be determined in a backward fashion by deciding all the $\theta_{t}$'s, from $\theta_{T-1}=\mathds{E}[w_{T}]=\mathds{E}[p(t)]$ to $\theta_{0}=\mathds{E}[w_{1}]$, all using Eq. (\ref{star}). This simple threshold control policy is our expected threshold algorithm (ETA) to treat online storage control.

\subsection{Optimality of ETA}
The optimality of ETA comes from the fact that the decision making at each time slot is a binary choice. With this fact, we can prove the following theorem:

\vspace{0.1cm}
\noindent \emph{Theorem}: ETA is the optimized storage control policy for the one-shot load serving problem.
\vspace{0.1cm}

\noindent \textbf{Proof}: We can prove this theorem by backward induction. Note that at $\tau=T-1$, $\theta_{T-1} = \mathds{E}[w_{T}]=\mathds{E}[p(T)]$ is optimal choice. This constructs the induction basis. For the induction part, it suffices to identify that our proposed optimal threshold $\theta_{t}$ is the solution to the first order optimality condition of (\ref{star}).

\vspace{0.1cm}
\noindent \textbf{Remark}: We have only proved that ETA is the optimal control policy for the one-shot load serving problem. To show it is the optimal control policy, we need show the one-shot load decomposition maintains the structure of the solution space between the original optimization problem and the decomposed optimization problem. This is our next step towards better understanding the problem structure.

\section{Data-driven Distribution Estimator}
\label{sec: DEAT}
In design the optimal online control policy, we assume the knowledge of exact distribution of $p(t)$'s. In this section, we relax this assumption and only assume the knowledge of type of distribution $f(p|\lambda)$ without the knowledge of parameters $\lambda$. This motivates us to design a data-driven distribution estimator to better project the distribution parameters.

\subsection{General Data-driven Estimator}
One straightforward way to conduct parameter estimation is to utilize the maximum likelihood method (MLE) \cite{myung2003tutorial}. We use MLE \footnote{We want to emphasize that MLE may not be the optimal distribution estimator (e.g., minimizing KL divergence may be a better choice for some cases, see \cite{anderson2004model} for a detailed discussion).} as an example to illustrate the incorporation process as new price data is available. 

Suppose we have collected realized price data $p = (p_{1},...,p_{t})$ and know these prices obey distribution $f(p|\lambda)$ with unknown parameters $\lambda$. To estimate the distribution parameters $\lambda$, we define the likelihood function $L(\lambda)$ as follows: 
\begin{equation}
\begin{aligned}
 L(\lambda) = f(p_{1}|\lambda)f(p_{2}|\lambda)\cdots f(p_{t}|\lambda).
\end{aligned}
\end{equation}

The MLE estimator maximizes the log-likelihood function, $\ln L(\lambda)$. The optimal parameter $\lambda^{*}$ given by MLE estimator need satisfy the following two conditions:
\begin{equation}
\begin{aligned}
\frac{\partial \ln L(\lambda)}{\partial \lambda_{i}}=0,&\quad&\forall{i},
\end{aligned}
\label{max1}
\end{equation}
and
\begin{equation}
\begin{aligned}
\frac{\partial^{2} \ln L(\lambda)}{\partial \lambda_{i}^{2}}<0,&\quad&\forall{i}.
\end{aligned}
\label{max2}
\end{equation}

\subsection{Data-driven Expected Threshold Algorithm}
For our optimal control policy design, at each time slot $t$, witnessing another new instance of dynamic price, $p(t)$, we can update the likelihood function as follows:
\begin{equation}
L(\lambda) \leftarrow L(\lambda)f(p(t)|\lambda).
\end{equation}
This will lead to a new estimation of distribution parameter, which need be utilized when deriving the optimal threshold as we do in ETA. We refer to such adaptive policy as the Data-driven Expected Threshold Algorithm (DETA). 

We want to emphasize that there is an initialization period for DETA when we only observe the prices and serve the demand without the help of storage system.

\section{Simulation Studies}
\label{sec:SimulationStudy}
We have proved that ETA is the optimal online control policy for one-shot load serving problem. However, comparing with the offline optimal benchmark, there are still differences. In this section, we start by evaluating ETA's regret ratio for one-shot load serving. Then, we observe that ETA's competitive ratio becomes stable in numerical studies. We further our understanding towards the risk of unknown by numerical studies.

\subsection{Dataset Characterization}
Dynamic pricing has not been widely adopted for residential users. Hence, we choose to use hourly real-time pricing data to characterize the stochastic nature in dynamic price. The data is collected from PJM during August, 2019 \cite{pricedata}. We use three distributions to approximate the price histogram: uniform distribution, half-normal distribution (light tail), and log-normal distribution (heavy tail). While log-normal distribution seems better fit the histogram, we select these distributions for better illustration of the tail performance. We randomly sample a period from the PJM users' load data for numerical study. The load data is collected in AEP area during August, 2019 \cite{demanddata}. We assume the users are equipped with storage devices with capacity that is $10\%$ of their peak demand. 

We conduct the simulation using the fitted price distributions. This eliminates all the possible price correlation in the real data. We intend to relax the \textit{i.i.d.} assumption on the price distribution in our future work.


\begin{figure*}[!t]
\centering
\subfloat[Uniform distribution]{\includegraphics[width=2.2in]{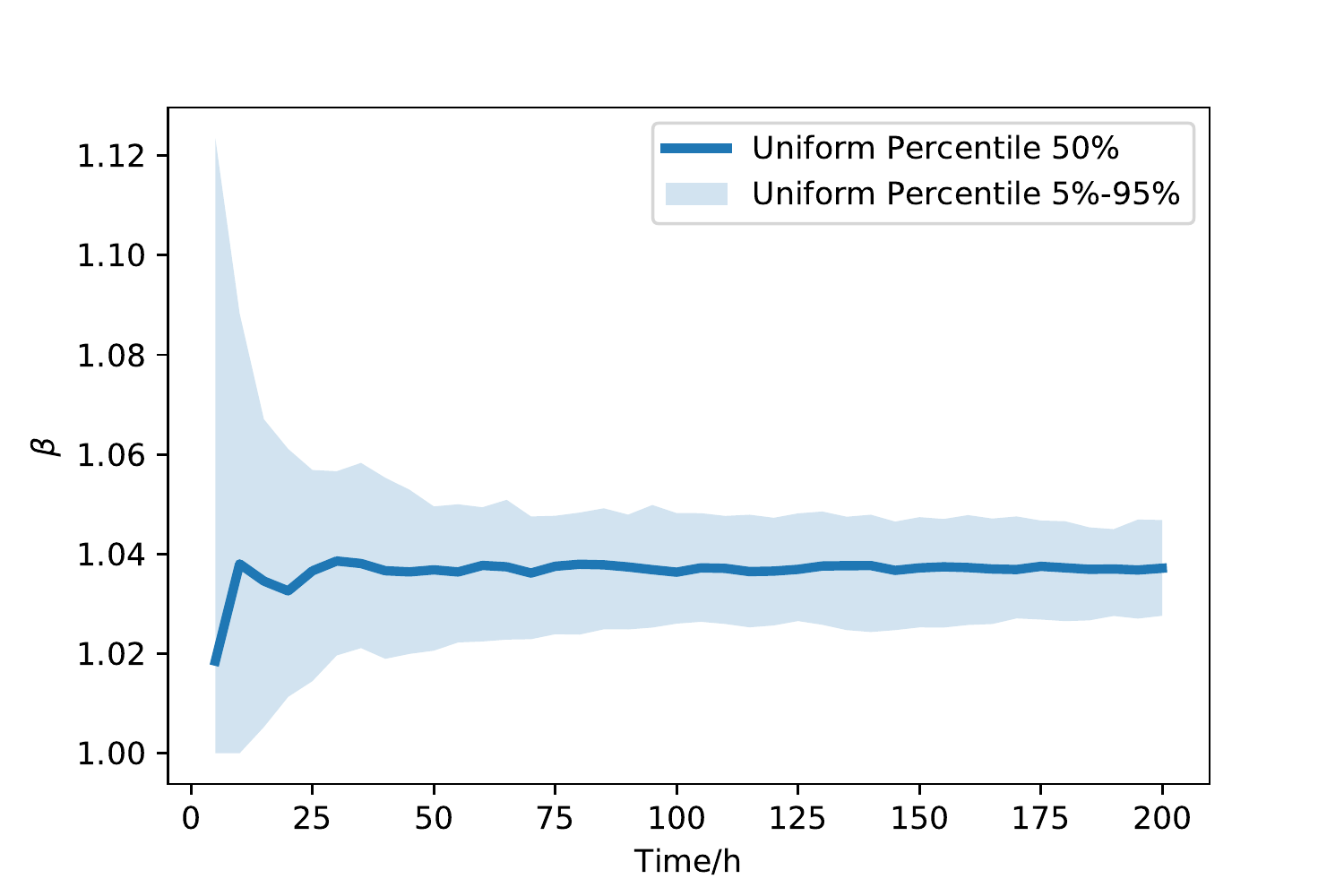}}
\hfil
\subfloat[Half-normal distribution]{\includegraphics[width=2.2in]{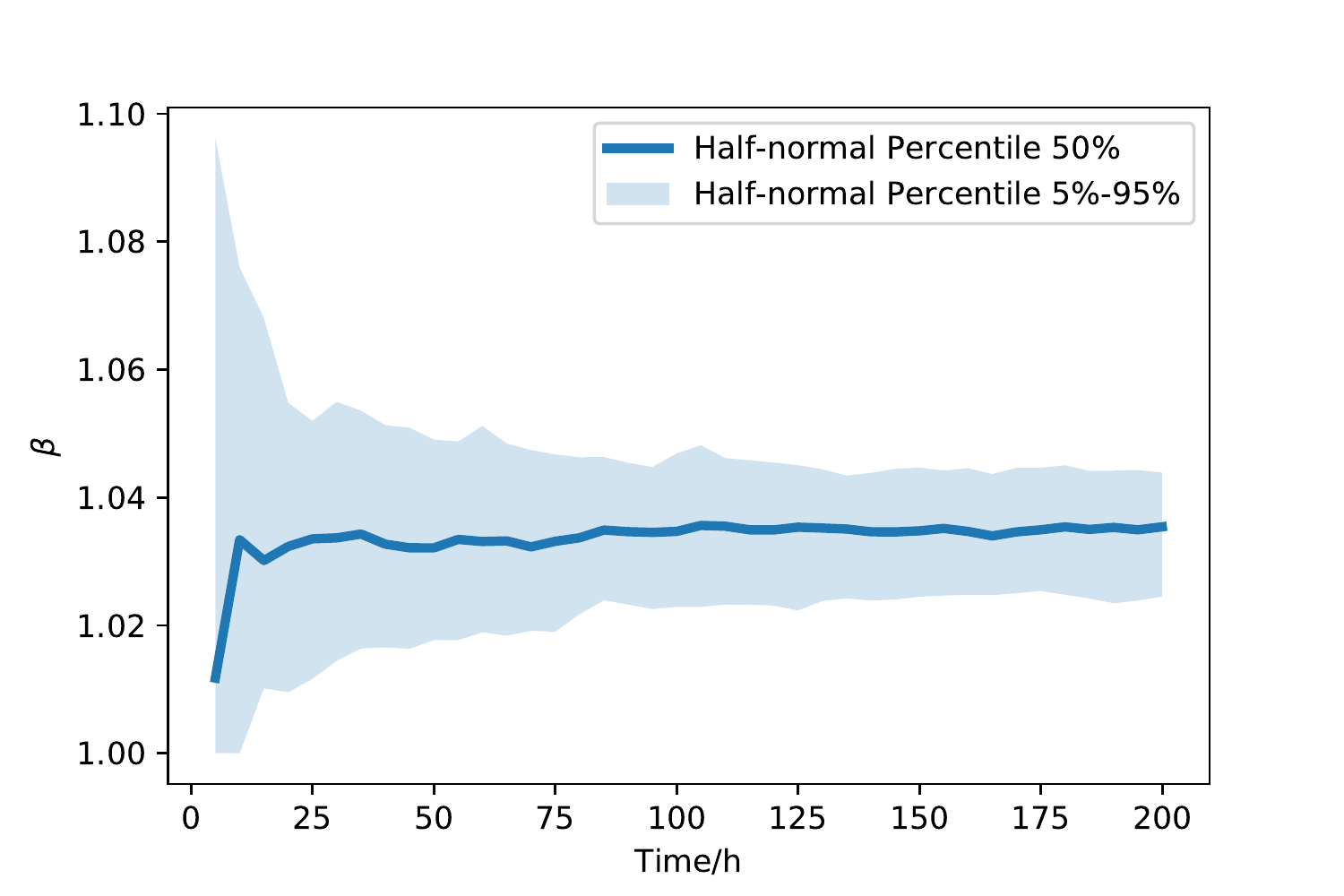}}
\hfil
\subfloat[Log-normal distribution]{\includegraphics[width=2.2in]{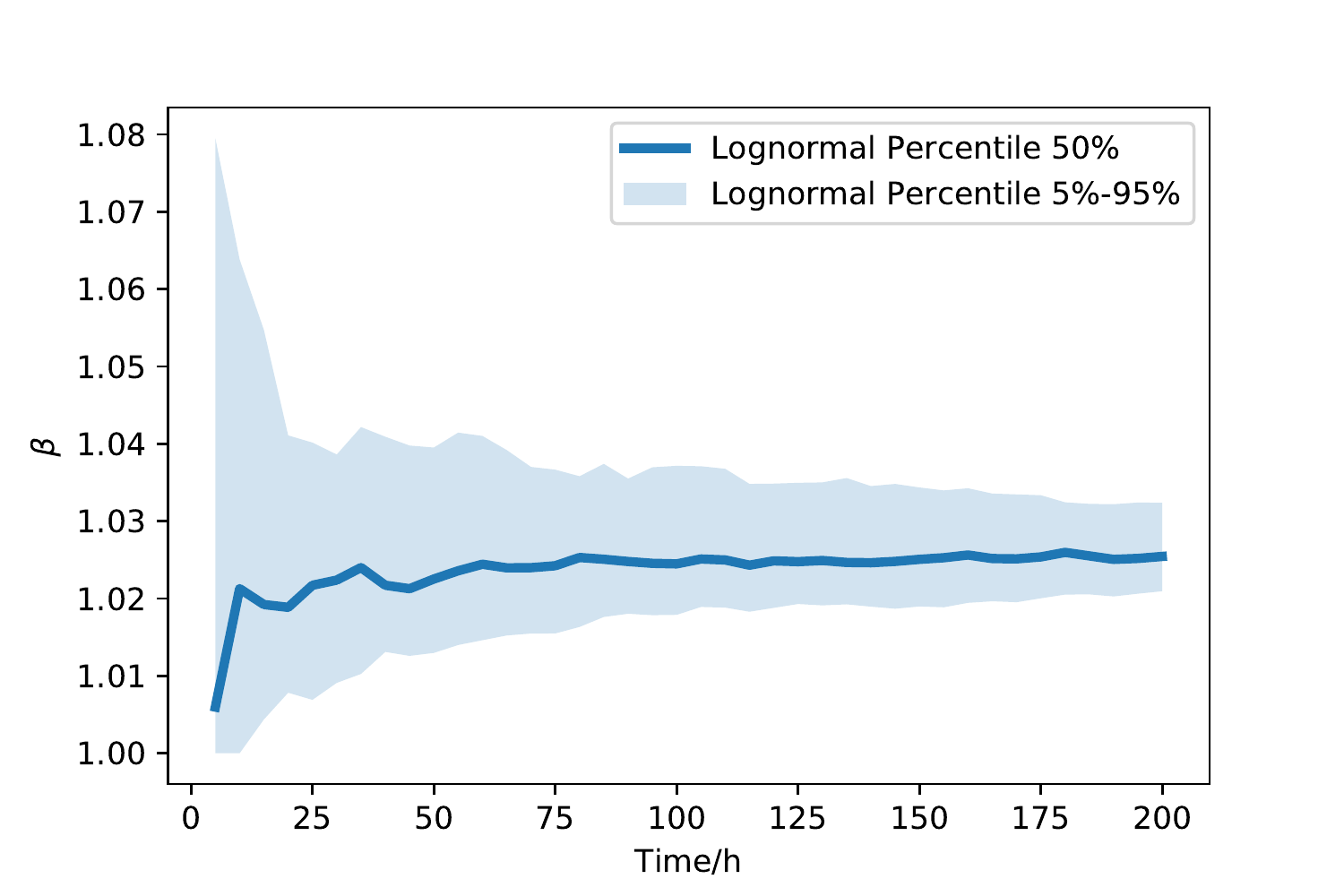}}
\caption{Competitive ratio for load serving.}\vspace{-0.2cm}
\label{fig 6}
\end{figure*}

\begin{figure*}[!t]
\centering
\subfloat[Uniform distribution]{\includegraphics[width=2.2in]{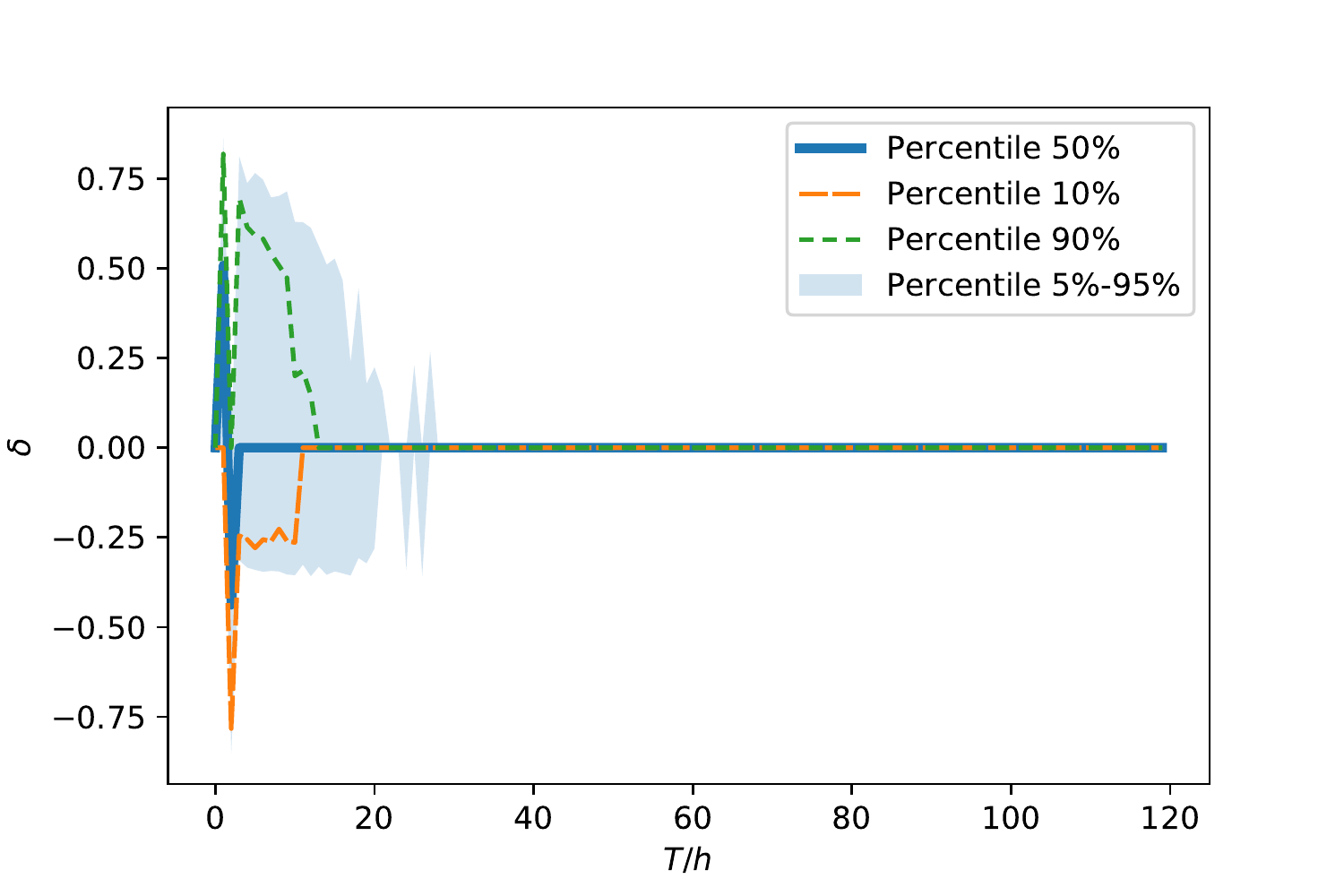}}
\hfil
\subfloat[Half-normal distribution]{\includegraphics[width=2.2in]{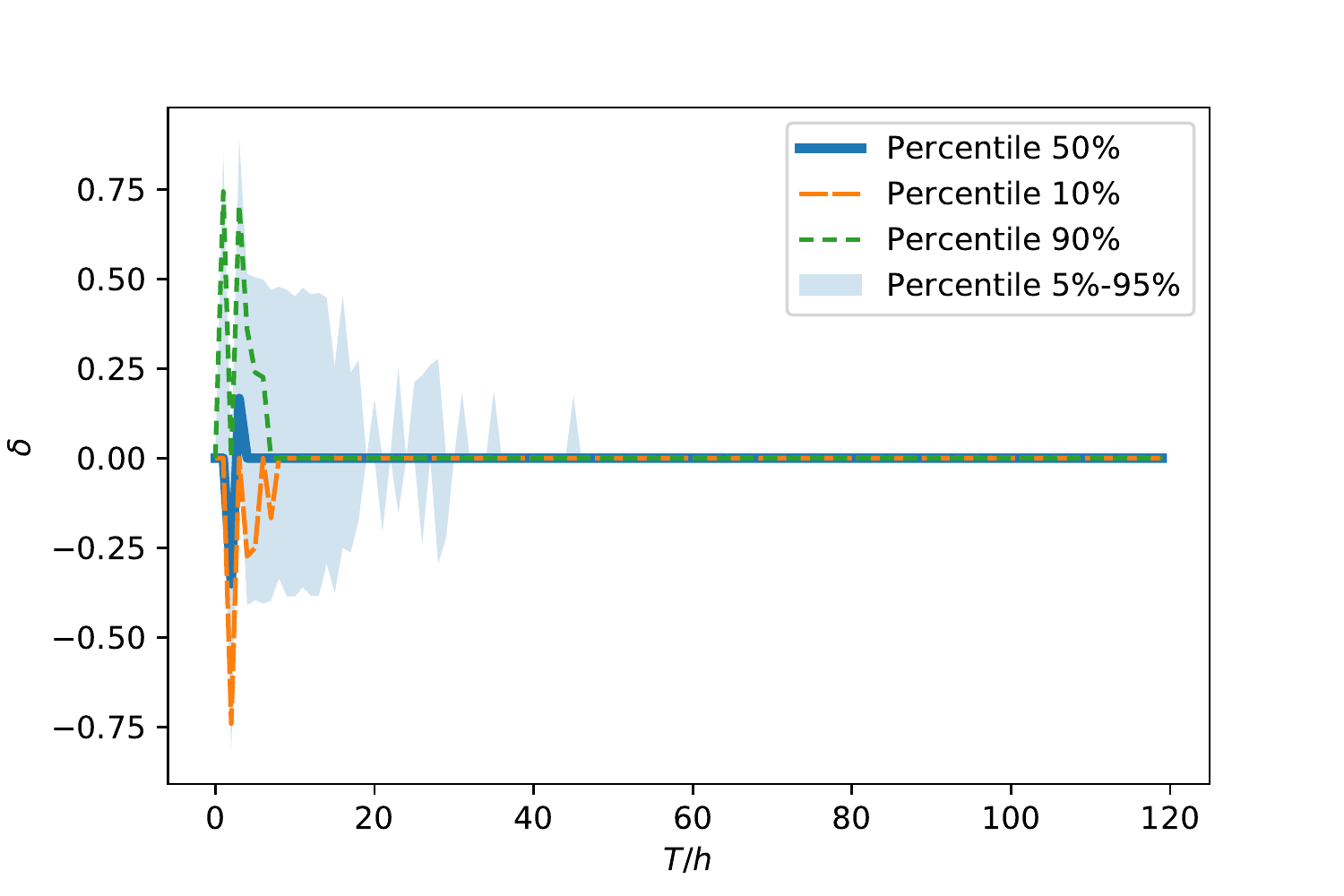}}
\hfil
\subfloat[Log-normal distribution]{\includegraphics[width=2.2in]{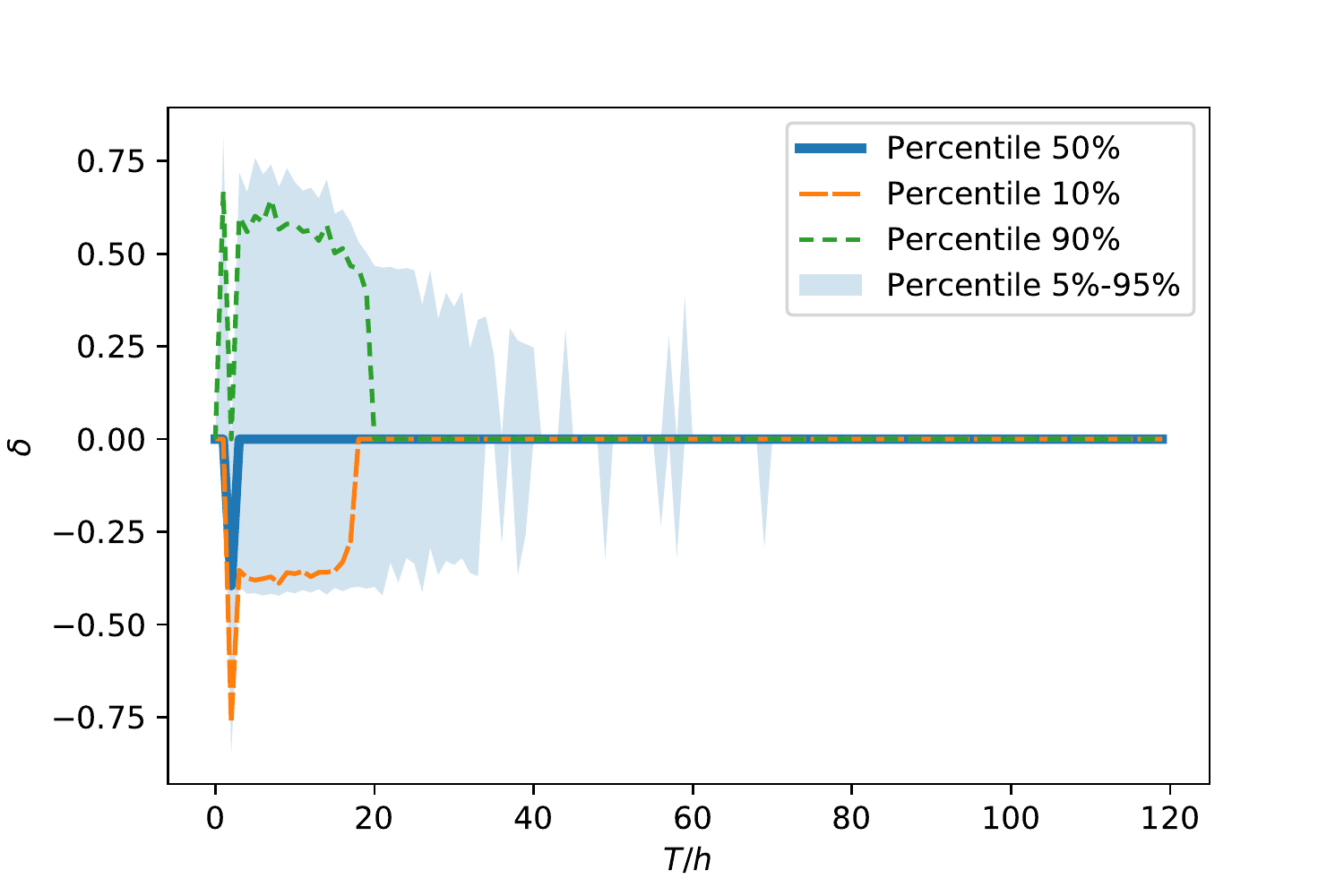}}
\caption{Evolution of $\delta$ with accumulated data.}\vspace{-0.3cm}
\label{fig 7}
\end{figure*}

\subsection{Evaluate ETA for One-shot Load Serving}
We first evaluate the performance of ETA for the one-shot load serving problem. To compare our ETA performance and offline optimal, we define regret ratio $\gamma$ as follows:
\begin{equation}
\gamma = \frac{cost(ETA)-OPT}{OPT}.
\end{equation}
where $cost(ETA)$ denotes the cost of serving the one-shot by ETA while OPT denotes the offline minimal cost. Figure. \ref{fig 5} plot the performance of ETA using the three fitted distributions. All cases display diminishing regret ratio, which implies they converge to the offline optimal rather fast. This illustrates the robustness of ETA to different price distributions.

\subsection{Evaluate ETA for Load Serving}
Next, we evaluate the performance of ETA for general load serving purposes. We define the competitive ratio $\beta$:
\begin{equation}
\beta = \frac{cost(ETA)}{OPT},
\end{equation}
where $cost(ETA)$ denotes the \textit{total} cost of ETA in serving load during certain period of time, while OPT denotes the corresponding offline minimal \textit{total} cost. Figure \ref{fig 6} plots the competitive ratio for the three fitted distributions. We observe that the competitive ratio becomes stable as time goes by. In all the three cases, the mean of $\beta$ is bounded by $1.04$.

\subsection{Examine the Risk of Unknown}
It's commonly believed that more data implies better performance. In this section, by comparing the performance between ETA and DETA, we examine the conventional wisdom for online storage control.

The implementation of DETA requires historical data, which helps decide the distribution parameters. We study the performance difference between ETA and DETA, denoted by $\delta$ , as time goes by (and hence, as data accumulate!):
\begin{equation}
\delta = cost(DETA)-cost(ETA).
\end{equation}

Figure \ref{fig 7} compares the evolution of $\delta$ for three fitted distributions. The $90\%$ percentile shows that in most cases, DETA converges to ETA extremely fast. In fact, for storage online control purpose, it suffices to observe tens of data for reasonably good performance.

\section{Conclusion}
\label{sec:con}
In this paper, we investigate the online storage control policy design facing dynamic prices. Based on one-shot load decomposition, we propose two control policies: ETA and DETA. We compare the performance of these two online control policies with the offline optimal, which demonstrates the robustness of the two policies to different price distributions.

However, much remains unknown. From theoretical point of view, we plan to investigate the theoretical bound for ETA's regret ratio for one-shot load serving problem. It is also interesting to examine how the load decomposition technique changes the structure of the optimization problem. More practically, we intend to relax the two assumptions. We are interested in relaxing the assumption of perfect load prediction and studying the impact of coupled uncertainties (from both dynamic price and the demand) on optimal storage control. We also would like to relax the \textit{i.i.d.} assumption of price distributions. This may yield a more practical control policy.

\bibliographystyle{ieeetr}
\bibliography{ref}

\end{document}